\documentclass{iaus}
\usepackage{graphicx}

\title[The DS/DT Project] 
{Time-Resolved Spectroscopy with SDSS}

\author[Bickerton et al.]   
{Steven Bickerton$^1$,
Carles Badenes$^2$,
Thomas Hettinger$^3$,
Timothy Beers$^3$,
\and Sonya Huang$^1$}

\affiliation{
  $^1$Dept. of Astrophysical Sciences, Princeton University, Princeton NJ, USA \\ 
  email: {\tt bick@astro.princeton.edu}\\[\affilskip]
  $^2$Dept of Physics and Astronomy, University of Pittsburgh, Pittsburgh PA, USA \\
  email: {\tt badenes@pitt.edu}\\[\affilskip]
  $^3$Dept of Physics and Astronomy and JINA: Joint Institute for Nuclear Astrophysics, \\
  Michigan State University, East Lansing MI, USA\\
  email: {\tt hettin12@msu.edu, beers@pa.msu.edu}\\[\affilskip]
}

\pubyear{2011}
\volume{285}  

\jname{New Horizons in Time Domain Astronomy}
\editors{R.E.M. Griffin, R.J. Hanisch, \& R. Seaman}
\begin{document}

\maketitle

\begin{abstract}
  We present a brief technical outline of the newly-formed `Detection
  of Spectroscopic Differences over Time' (DS/DT) project.  Our
  collaboration is using the individual exposures from the SDSS
  spectroscopic archive to produce a uniformly-processed set of
  time-resolved spectra.  Here we provide an overview of the
  properties and processing of the available data, and highlight the
  wide range of time baselines present in the archive.

\end{abstract}

\firstsection 
\section{Introduction}

The Sloan Digital Sky Survey (SDSS) \cite{york00} has been in
operation for over a decade, and through three separate phases (SDSS
I, II, and III) has accumulated an extensive archive of photometric
and spectroscopic astrophysical observations.  Here we describe a
newly-formed data mining collaboration, the `Detection of
Spectroscopic Differences over Time' or DS/DT project, to search for
variability in the SDSS spectra.

Though cadences and exposure times varied, the SDSS spectra were
generally observed in three or more exposures, with exposure times of
at least 900s.  These individual exposures were combined to
produce the final spectra released to the community.

Only a handful of groups have examined the sub-spectra for evidence of
variability: \cite{hilton10} examined flaring in M-dwarfs, while
\cite{mullally09} and \cite{badenes09} used radial velocity shifts to
identify WD-WD binaries.  Radial velocities have also been used by
\cite{rebassa10} to identify binaries.  With such an enormous
diversity of objects, from QSOs and AGNs, to stars of all spectral
types; there is a vast amount of time variable data which has never
been evaluated.

The primary objective of our work is to produce a uniformly-processed
archive of the individual sub-spectra for all SDSS spectroscopic
observations.  We are developing a data mining pipeline targeting some 
specific forms of variability (radial velocities, flaring, etc), and
also serendipitous anomalous variability.

\section{Data Processing}
\label{sec:processing}

Our pipeline is an extension of the SDSS spectroscopic pipeline, {\tt
  spectro}, used to produce seventh data release of the Sloan Digital
Sky Survey \cite{abazajian09}.  The original {\tt spectro} pipe
consists of three principle stages for (1) extraction and calibration
of the raw spectroscopic traces, (2) stacking of the sub-spectra and
stitching of the data from the red and blue cameras, and (3) object
classification and redshift determination.  We have written a
modified 2nd stage to perform only the red/blue stitching for the
individual exposures.  Our new combine step generates individual
per-object output files containing the full co-added spectra and their
associated sub-spectra.

\section{Time Baselines}
\label{sec:baselines}

\begin{figure}[h]
\begin{center}
 \includegraphics[width=5.0in]{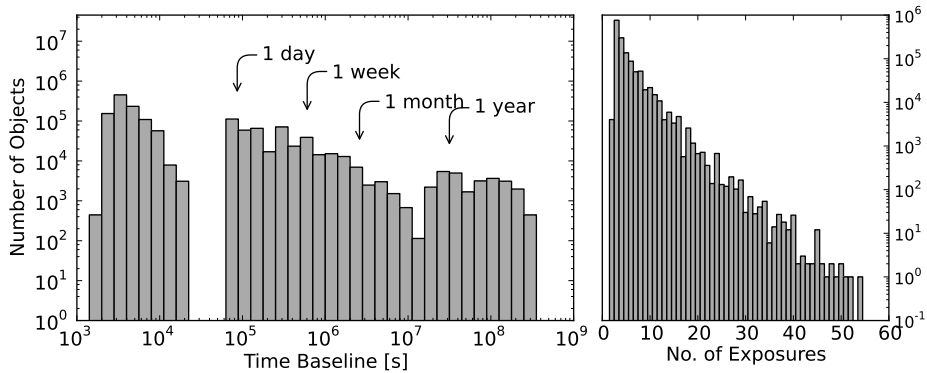} 
 \caption{The per-object time baselines sampled by the SDSS
   sub-spectra (left), and the number of targets having a given number
   of exposures available (right). }
   \label{fig:baselines}
\end{center}
\end{figure}

During the SDSS observing program, a wide range of time baselines were
sampled.  A given plate (640 fibers) was typically observed in three
back-to-back 900s exposures, and in most cases the baseline is
therefore $\lesssim$30 minutes.  However, plates which were incomplete
at the end of an observing session were continued on the following
night, giving many plates a baseline of $>$12 hours.  In some cases,
previously-observed plates were later replugged for further
observation, providing baselines of several weeks or sometimes much
longer.  Finally, some targets were observed on multiple plates, again
providing baselines from weeks to months.  Figure~\ref{fig:baselines}
shows histograms of the time baselines sampled (measured between the
mid-points of the first and last available exposures) and the number of
exposures.  The sample includes the SDSS I and II spectra, and the
SEGUE-II subset of the SDSS III spectra.

\section{Summary}
\label{sec:summary}

The DS/DT project is a newly-formed data mining project to explore
time variability in the sub-spectra of the SDSS spectroscopic archive.
We have developed a branch pipeline to process and collate the
sub-spectra for each target into individual FITS files.  It is our
intention to make our full data set available.  Those interested in
testing and providing feedback on our preliminary data are invited to
contact us.


\end{document}